\documentclass[letterpaper,aps,prc,superscriptaddress,nofootinbib,showpacs,floatfix]{revtex4-1}
\usepackage[utf8]{inputenc}
\usepackage{amsmath}
\usepackage{amsfonts}
\usepackage{amssymb}
\usepackage{amsthm}
\usepackage{bm}
\usepackage{hyperref}
\usepackage[english]{babel}
\usepackage{fontenc}
\usepackage{graphicx}
\usepackage[margin=1in]{geometry}
\usepackage{enumerate}
\usepackage{textcomp}
\usepackage[toc,page]{appendix}
\usepackage{slashed}
\usepackage{color}

\usepackage{setspace}
\graphicspath{ {figure/}}

\hypersetup{colorlinks=true, linkcolor=blue, citecolor=blue, urlcolor=blue}

{\large }

\begin{document}
\title{Effects of  Non-Vanishing Net Charge in Balance Functions}

\author{ Claude~Pruneau }
\email{claude.pruneau@wayne.edu}
\affiliation{Department of Physics and Astronomy, Wayne State University, Detroit, 48201, USA}
\author{ Victor~Gonzalez }
\email{victor.gonzalez@cern.ch}
\affiliation{Department of Physics and Astronomy, Wayne State University, Detroit, 48201, USA}
\author{ Brian~Hanley }
\email{bghanley@wayne.edu}
\affiliation{Department of Physics and Astronomy, Wayne State University, Detroit, 48201, USA}
\author{Ana Marin} 
\email{a.marin@gsi.de}
\affiliation{GSI Helmholtzzentrum f\"ur Schwerionenforschung, Research Division and ExtreMe Matter Institute EMMI, Darmstadt, Germany}
\author{ Sumit~Basu }
\email{sumit.basu@cern.ch}
\affiliation{Lund University, Department of Physics, Division of Particle Physics, Box 118, SE-221 00, Lund, Sweden}
\date{Sept 2022}

\begin{abstract}
 We investigate the impact  of non-vanishing net-charge in collision systems on measurements of  balance functions and their integrals. We  show that the nominal balance function definition yields integrals that deviate from unity because of the  non-vanishing net-charge. However, the  integral of unified balance functions is shown to appropriately converge to unity when measured in a sufficiently wide experimental  acceptance.  We furthermore explore the rate of convergence of unified balance functions integrals and study distortions imparted on the shape of balance functions when measurements are carried out  in limited transverse momentum ($p_{\rm T}$) and rapidity ($y$) acceptances, such as those featured by experiments at  the Large Hadron Collider or the Relativistic Heavy Ion Collider. 
We show that the shape and integral of unified balance functions may be strongly biased  by reductions in the rapidity and transverse momentum acceptances of existing experiments. 
 
\end{abstract}

\maketitle

\section{Introduction}
\label{sec:introduction}

Balance functions (BFs) were initially proposed as a tool to investigate the evolution of particle production in heavy-ion collisions~\cite{Bass:2000az,Pratt:2002BFLH,Jeon:2002BFCF} and were later found to have sensitivity to  various  properties of the quark gluon plasma (QGP) matter  produced in these collisions such as  the  diffusivity of light quarks and QGP susceptibilties~\cite{S.GavinAPHA:2006Diffusion,Jeon:2002BFCF,Pratt:2019pnd}. The 
former brings about  broadening of azimuthal balance functions, particularly those of heavy particle pairs, and the latter determines the relative yields of charge (strangeness or baryon number) balancing partners~\cite{Pratt:2019pnd, Pratt:2015jsa, Pratt:2021xvg, Pruneau:2007ua}, while the longitudinal dependence of balance functions is sensitive to the system's evolution~\cite{Pratt:2002BFLH,ALICE:2021hjb,Basu:2020ldt}. Measurements of balance functions and their integrals are thus of great interest. The  shape and integral of BFs are however sensitive to a number of physical and instrumental effects~\cite{Bialas:2004BFCO,Pratt:2003inv,Pratt:2022xbk,Wang:2011za}. It is thus important to quantitatively understand 
the impact of such effects on the shape and integral of BFs. 

The main two  goals of this work are (1)  to  investigate the impact  of non-vanishing net-charge and (2) study the effects of  finite acceptance on measurements of general balance functions (GBFs). We first show that the nominal GBF definition~\cite{Pratt:2002BFLH,Jeon:2002BFCF} yields integrals that are determined by the non-vanishing net-charge of colliding systems, whereas the integral of unified balance functions (UBFs), introduced in Ref.~\cite{Pruneau:2022gvt}, appropriately  converge to unity when measured in a sufficiently wide experimental  acceptance. The presence of non-vanishing net-charge in collision systems can thus be compensated for, in principle,  with the use of UBFs when measurements are carried out in full or nearly full kinematic acceptances. We explore the rate of convergence of UBF integrals and distortions imparted on the shape of BFs when measurements are carried out  in limited transverse momentum ($p_{\rm T}$) and rapidity ($y$) acceptances, such as those    
featured by experiments at  the Large Hadron Collider (LHC) or the Relativistic Heavy Ion Collider (RHIC). 
We also show that the shape and integral of balance functions are unfortunately very much affected by reductions in the rapidity and transverse momentum acceptances matching current experimental capabilities. 
To accomplish the above goal, we  use simulations of pp collisions performed with the PYTHIA8 model based on the assumption that it provides a reasonable quantitative  account of particle production at ultra relativistic energy.   

Integral of BFs are closely related to the magnitude of second order  net-charge cumulants, $\kappa_2$, nominally of interest in the determination of QGP susceptibilties at LHC and RHIC~\cite{Pruneau:2019baa}, as well as searches for the critical point of nuclear matter based on the beam energy scan (BES) at RHIC~\cite{Braun-Munzinger:2019yxj,Rustamov:2022hdi}. 
The results reported in this work thus also have bearings on the limitations of such measurements carried out at those facilities. In particular, it should be clear that modification of the shape of BFs and their integrals caused by reduced kinematic acceptances shall then also have a somewhat arbitrary impact,  virtually impossible to  correct for,  on the magnitude of measured values of  $\kappa_2$ at RHIC and LHC by ongoing experiments. 

This paper is organized as follows: Section~\ref{sec:Simulations} presents a short discussion of the model and techniques used to carry out the simulations and analyses presented in this work. Section~\ref{sec:BF} presents simulations of GBFs obtained with the nominal definition of Pratt et al.~\cite{Bass:2000az,Pratt:2002BFLH,Jeon:2002BFCF} and articulates that integrals of these GBFs do not converge to unity for non-vanishing net-charge systems. The following section reports results obtained with the UBFs, introduced in Ref.~\cite{Pruneau:2022gvt}, whose integrals converge to unity in a sufficiently wide acceptance, whereas Sec.~\ref{sec:impact} discusses the impact of restricted measurement acceptances on the shape and integral of UBFs. A summary of this work is presented in Sec.~\ref{sec:summary}. 
 
\section{Simulation Model -- PYTHIA8}
\label{sec:Simulations}

\begin{figure}[!ht]
	\centering
	\includegraphics[width=0.5\linewidth,trim={5mm 14.5mm 2mm 2mm},clip]{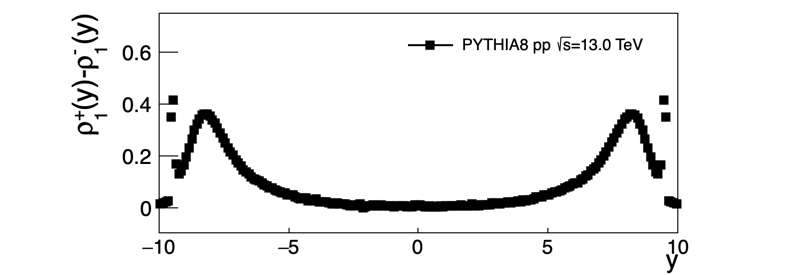}
	\includegraphics[width=0.5\linewidth,trim={5mm 0.5mm 2mm 4.2mm},clip]{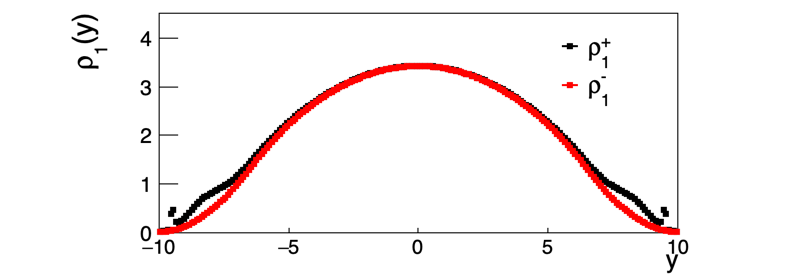}
	\caption{Bottom: Densities of positive particles, $\rho_1^+(y)$, and negative particles, $\rho_1^-(y)$, produced in $\sqrt{s}= 13.0$ TeV  pp collisions according to PYTHIA8 (MONASH) plotted as a function of the particles rapidity; Top: Density difference  $\rho_1^+(y)-\rho_1^-(y)$ vs. particle rapidity. }
	\label{fig:Pythia:Rho1VsY}
\end{figure}

The charge BFs of particles produced in 
high-energy proton-proton (pp) collisions are investigated theoretically based on simulations carried out with the PYTHIA8  Monte Carlo event generator operated with the MONASH 2013 tune~\cite{Skands:2014pea} with color reconnection.  PYTHIA8 is  based on a QCD  description of quark and gluon interactions    at leading order (LO) and uses the Lund string fragmentation model for 
high-$p_{\rm T}$ parton hadronization while the production of soft particles (i.e., the underlying event) is handled through fragmentation of mini-jets from initial and final state radiation, as well as multiple parton interactions~\cite{Sjostrand:2007gs}.  Studies reported in this work are carried with PYTHIA running in minimum-bias mode, with soft QCD processes and color reconnection turned ON.  Events are generated and analyzed on the fly, within a simple analysis framework~\cite{pruneauX},  to limit the use of large storage space. Analyses were run on the Wayne State computing grid in groups of 10 jobs, each with 10 sub-jobs, and 300,000 events per sub-job. This enabled efficient and rapid use of the grid: each run (e.g., pp collisions at a specific beam energy) were completed in less than two hours. The output of sub-jobs were summed to yield high-statistic single- and pair-densities, which were subsequently used to compute correlation functions and balance functions. The ten jobs were then combined and used to compute statistical uncertainties  on the amplitude of these functions  using the sub-sample method. 
PYTHIA8 has been highly successful in reproducing a wide variety of  measurements and observables in pp collisions ranging from the few GeV scale to the TeV scale ~\cite{Sjostrand:2007gs}.
As a first step  in the use of PYTHIA8 towards  studies of charge BFs, we explicitly verified that the MC event generator conserves charge on an event-by-event basis.  Indeed, integrating all particles in the range $-10 < y < 10$ for pp collisions ranging from $\sqrt{s} = 1$ TeV to 13 TeV, we find that the net-charge $Q \equiv N_1^+ - N_1^-$, computed event-by-event, is exactly equal to $2$, i.e., the sum of the (net) charge of the incoming beams. We also verified, based on Fig.~\ref{fig:Pythia:Rho1VsY}, which displays the densities  $\rho_1^+(y)$,  $\rho_1^-(y)$, and their difference, that the integral of these densities in the $-10 < y < 10$ range differs by exactly two units of charge. 
Note that the densities $\rho_1^+(y)$,  $\rho_1^-(y)$ feature narrow spikes at $\pm 9.4$   corresponding to protons at beam rapidities from single diffractive interactions~\cite{Jezabek:2021oxg}, protons forming  broad peaks at $\pm 8.2$ from elastic collisions, and extensive charged particle production over a broad range of  rapidities, $|y|<7$, in which only a small rapidity dependent  excess of positively charged particles  is produced. 

\section{Basic General Balance Functions }
\label{sec:BF}

We first examine BFs obtained with a basic definition similar to that first proposed by Bass et al.~\cite{Bass:2000az}.  Herewith,  the identity of  particle species is denoted  with Greek letters $\alpha$, $\beta$, etc, and their respective anti-particles with barred letters $\bar \alpha$, $\bar\beta$, etc. Single and pair densities of species $\alpha$ and $\beta$ are denoted 
$\rho_1^{\alpha}(\vec p_{\alpha})$ and $\rho_2^{\alpha\beta}(\vec p_{\alpha},\vec p_{\beta})$, respectively, where  $\vec p_{\alpha}$ and $\vec p_{\beta}$  represent the 3-momenta of the particles. Correlation functions can be computed/measured for particles produced within  specific, possibly distinct, acceptances denoted $\Omega_{\alpha}$. In practice, throughout this work, we simplify the calculations and assume all particles are measured over the full azimuth, $0\le \varphi < 2\pi$, within a specific transverse momentum range $p_{\rm T,\min} \le p_{\rm T} < p_{\rm T,\max}$, and a symmetric rapidity range $-y_0 < y < y_0$. Additionally, for brevity of notation,  only the rapidity dependence, $y_0$, is explicitly shown in equations. 

Differential GBFs  were introduced by Bass et al.  based on averages of  conditional (differential) densities $\rho_2^{\alpha|\beta}(y_1|y_2)$, i.e.,  corresponding to density of a species $\alpha$ at $y_1$ given a particle of species $\beta$ is detected at $y_2$~\cite{Bass:2000az}. Following  Ref.~\cite{Pruneau:2022gvt}, however, we opt for a formulation involving two distinct functions  $B^{\alpha|\bar\beta}(y_1|y_2)$ and $B^{\bar\alpha|\beta}(y_1|y_2)$  defined as
\begin{align}
	\label{eq:B2PrattA}
    B^{\alpha|\bar\beta}(y_1|y_2) &= \frac{1}{\langle \rho_1^{\bar\beta}\rangle(y_2)}\left[ \rho_2^{\alpha|\bar\beta}(y_1|y_2) - \rho_2^{\bar\alpha|\bar\beta}(y_1|y_2) \right], \\
	\label{eq:B2PrattB}
    B^{\bar\alpha|\beta}(y_1|y_2) &= \frac{1}{\langle \rho_1^{\beta}\rangle(y_2)}\left[ \rho_2^{\bar\alpha|\beta}(y_1|y_2) - \rho_2^{\alpha|\beta}(y_1|y_2) \right],
\end{align}
where $y_2$ is the rapidity of the reference particle of species $\beta$ ($\bar\beta$) and $y_1$ is the rapidity of the particle of species $\bar\alpha$ ($\alpha$) balancing the charge.  Experimentally, as discussed in detail in Ref.~\cite{Pruneau:2022gvt}, one must consider averages of $B^{\alpha|\bar\beta}(y_1|y_2)$ and of $B^{\bar\alpha\beta}(y_1|y_2)$ over the rapidity $y_2$ of the reference particle. This leads to bounded balance functions computed according to 
\begin{align}
	\label{eq:B2vsy1y2}
    B^{\alpha\bar\beta}_{Q=0}(y_1,y_2|y_0) &= \frac{1}{\langle N_1^{\bar\beta}\rangle(y_0)}\left[ \rho_2^{\alpha\bar\beta}(y_1,y_2) - \rho_2^{\bar\alpha\bar\beta}(y_1,y_2) \right], \\
	\label{eq:B2vsy1y2Bis}
    B^{\bar\alpha\beta}_{Q=0}(y_1,y_2|y_0) &= \frac{1}{\langle N_1^{\beta}\rangle(y_0)}\left[ \rho_2^{\bar\alpha\beta}(y_1,y_2) - \rho_2^{\alpha\beta}(y_1,y_2) \right],
\end{align}
where $\langle N_1^{\bar\beta}\rangle(y_0)$ and $\langle N_1^{\beta}\rangle(y_0)$ are respectively  the event-ensemble averages of the yield of  particles of species $\bar\beta$ and  $\beta$  in the acceptance $|y| < y_0$, and the label $Q=0$ indicates these expressions are computed assuming the net-charge of the system is vanishing. We shall evidently also consider the arithmetic average 
\begin{align}
\label{eq:Bs}
B^{\rm s}\equiv (B^{\alpha\bar\beta} + B^{\bar\alpha\beta})/2,
\end{align}
which essentially corresponds to the  original definition of Bass et al.~\cite{Bass:2000az}. However, it should be noted that splitting the original BF definition  into two independent functions $B^{\alpha\bar\beta}$ and $B^{\bar\alpha\beta}$ provides the additional advantage, nominally, of enabling independent investigations of  the yield of a negative particles of type ``$\alpha$" at rapidity $y_{alpha}$ when a positive particle of type ``$\beta$" is observed at rapidity $y_{\beta}$, and conversely, the yield of a positive particles of type ``$\alpha$" at rapidity $y_{alpha}$ when a negative particle of type ``$\beta$" is observed at rapidity $y_{\beta}$. Original BFs were not designed to be split into two pieces. The split functions considered in this work thus provide new and distinct information that is  potentially useful   towards the understanding of charged (as well as strange  or charmed)  particle production.  It should be additionally noted that while the functions $B^{\alpha\bar\beta}$ and $B^{\bar\alpha\beta}$ may prove to be rather similar at ultra high beam energy, they are likely to differ significantly in both magnitude and shape at low beam energy. Indeed, at high beam energy, particle production is dominated by pair creation processes that yield ratios of anti-particles and particles converging towards unity, but at low energy, ratios of anti-particle to particle yields   significantly  deviate from unity: the functions $B^{\alpha\bar\beta}$ and $B^{\bar\alpha\beta}$ shall then accordingly be also different and independently provide valuable information about particle production.

\begin{figure}[!ht]
	\includegraphics[width=0.32\linewidth,trim={8mm 11mm 24mm 3mm},clip]{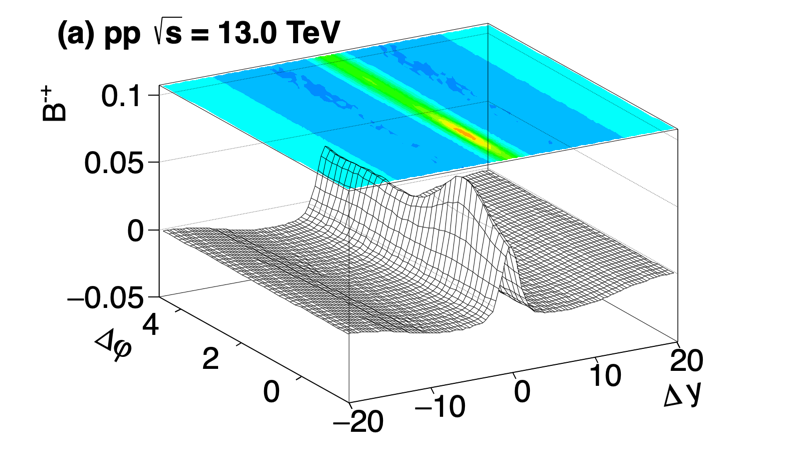}
	\includegraphics[width=0.32\linewidth,trim={8mm 11mm 24mm 3mm},clip]{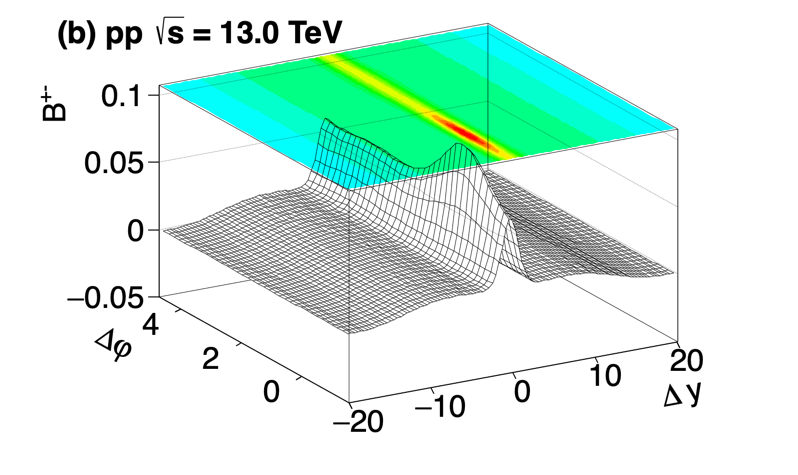}
	\includegraphics[width=0.32\linewidth,trim={8mm 11mm 24mm 3mm},clip]{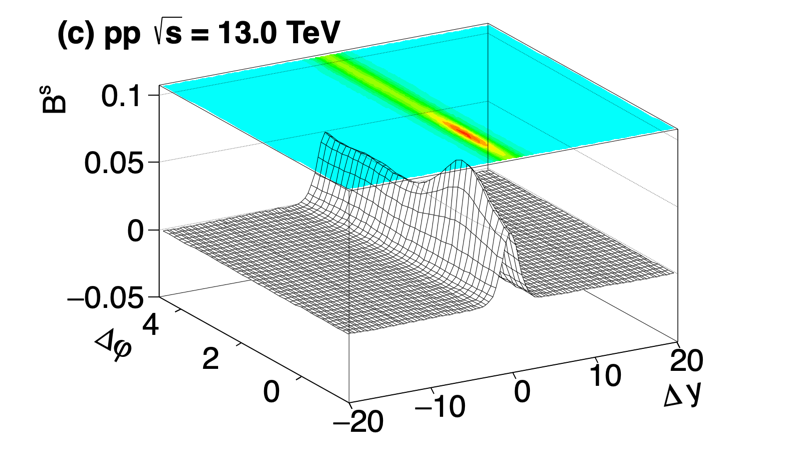}
	\caption{ Balance functions  (a) $B^{-+}$, (b) $B^{+-}$, and (c) $B^{\rm s}$ for charged particles with $p_{\rm T}>0$ calculated using Eqs.~(\ref{eq:B2vsy1y2},~\ref{eq:B2vsy1y2Bis},~\ref{eq:Bs})
	\label{fig:Pythia:PrattB2in2D} for pp collisions at $\sqrt{s}= 13$ TeV simulated with PYTHIA8.}
\end{figure}

Figure~\ref{fig:Pythia:PrattB2in2D} presents  balance functions $B^{-+}(\Delta y,\Delta\varphi)$,    $B^{+-}(\Delta y,\Delta\varphi)$, and their average $B^{\rm s}(\Delta y,\Delta\varphi)$ computed with Eqs.~(\ref{eq:B2vsy1y2},~\ref{eq:B2vsy1y2Bis},~\ref{eq:Bs}) for  pp collisions at $\sqrt{s}=13.0$ TeV
simulated with PYTHIA8.  Calculations were carried out using full azimuth, $p_{\rm T}>0$,   and  $-10 < y < 10$ to examine the full range of particle correlations.  First observe that $B^{-+}(\Delta y,\Delta\varphi)$ and $B^{+-}(\Delta y,\Delta\varphi)$ both feature a ``short-range" component extending roughly in the range $-5<\Delta y<5$ and  ``long-range" components extending all the way to $\Delta y\sim 2y_{\rm b}=\pm 19$. Also note that the short-range component of the BFs features an extended distribution in azimuth with a slight excess (or peak) on the near-side of the correlation function, i.e., centered at $\Delta y=0$, $\Delta \varphi = 0$.  

\begin{figure}[!ht]
\includegraphics[width=0.32\linewidth,trim={8mm 1mm 26mm 5mm},clip]{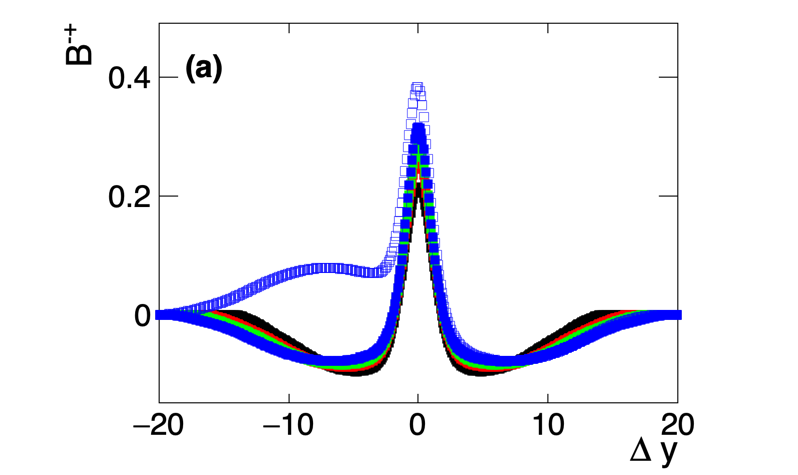}
\includegraphics[width=0.32\linewidth,trim={8mm 1mm 26mm 5mm},clip]{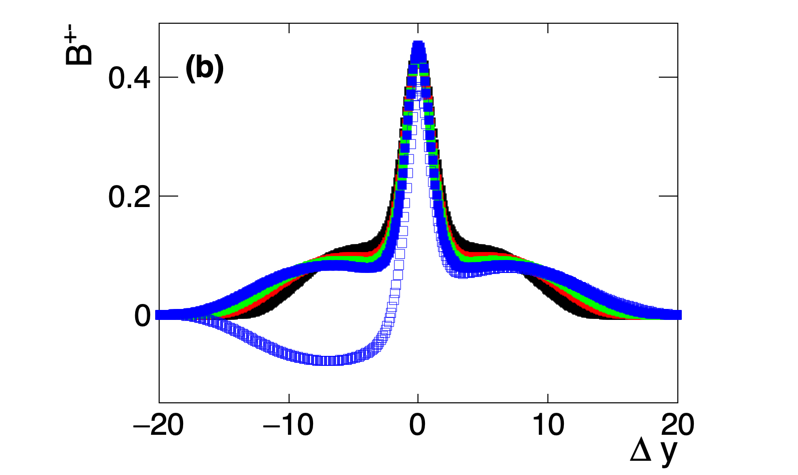}
\includegraphics[width=0.32\linewidth,trim={8mm 1mm 26mm 5mm},clip]{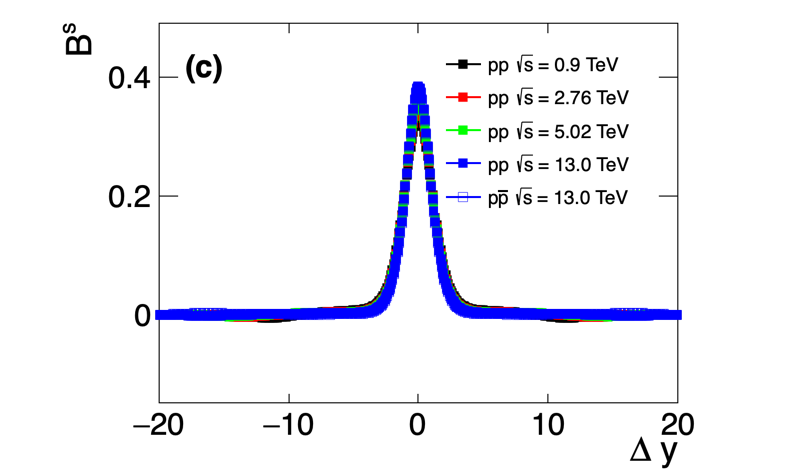}
\caption{ Projections onto the $\Delta y$ axis of balance functions (a)  $B^{-+}$, (b) $B^{+-}$, and (c) $B^{\rm s}$ calculated using Eqs.~(\ref{eq:B2vsy1y2},~\ref{eq:B2vsy1y2Bis},~\ref{eq:Bs}) for pp collisions at $\sqrt{s}= 0.9, 2.76, 5.02, 13$ TeV and $\rm p\bar p$ collisions at $\sqrt{s}= 13$ TeV simulated with PYTHIA8. }
\label{fig:BFvsy}
\end{figure}

The $\Delta y$ dependencies  of the short and long range components are easier to visualize in Fig.~\ref{fig:BFvsy} that displays projections of the BFs onto the rapidity $\Delta y$ axis. For illustrative purposes, this figure  includes projections of BFs of pp collisions at $\sqrt{s}=$ 0.9, 2.76. 5.02, and 13 TeV as well as $\rm p\bar p$ collisions at  $\sqrt{s}=13$ TeV.  
The short-range components $B^{-+}(\Delta y,\Delta\varphi)$ and $B^{+-}(\Delta y,\Delta\varphi)$ are  of similar strength and shape for all energies considered as well as for $\rm p\bar p$ collisions. However,  projections of BFs onto $\Delta y$  also  exhibit strong long-range components extending to twice the beam rapidity. The presence of these long range components is easy to understand: creating particle pairs at central rapidity ($y<5)$ requires beam protons lose kinetic energy thereby yielding particles spread across a large range of rapidities. These rapidity shifted particles are correlated to created pairs and thus appear prominently in $B^{-+}(\Delta y,\Delta\varphi)$ and $B^{+-}(\Delta y,\Delta\varphi)$ as long range components. We find, additionally, that these long-range components have similar shape but different signs  in $B^{-+}(\Delta y,\Delta\varphi)$ and $B^{+-}(\Delta y,\Delta\varphi)$: they identically vanish in the average $B^{\rm s}$ shown in Fig.~\ref{fig:BFvsy} (c).

While long range correlations  may be of interest from the perspective of beam stopping studies, e.g., what fraction of the incoming proton momentum is lost and converted into produced particles, their observation in experiments poses a considerable challenge given the very small angles (relative to the beam direction) required for their detection. They are also of limited interest to 
understand particle production at central rapidity. 
\begin{figure}[!ht]
	\includegraphics[width=0.32\linewidth,trim={8mm 1mm 26mm 3mm},clip]{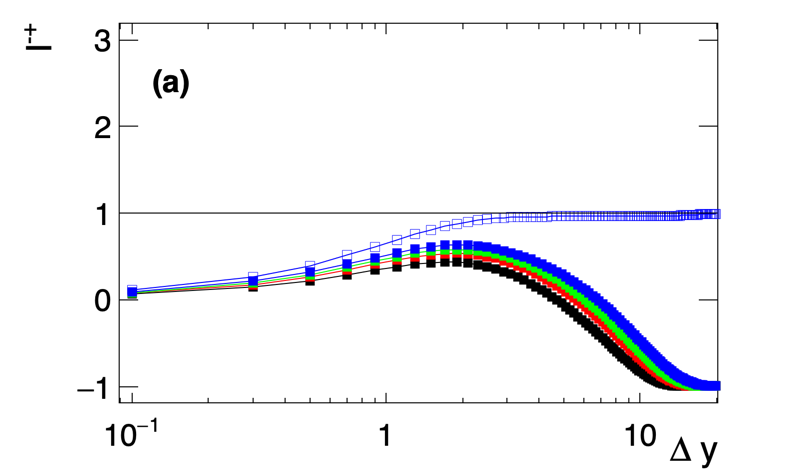}
	\includegraphics[width=0.32\linewidth,trim={8mm 1mm 26mm 3mm},clip]{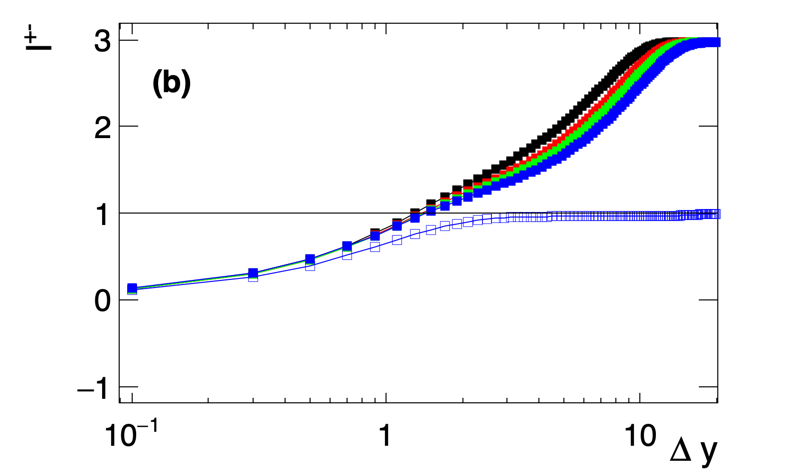}
	\includegraphics[width=0.32\linewidth,trim={8mm 1mm 26mm 3mm},clip]{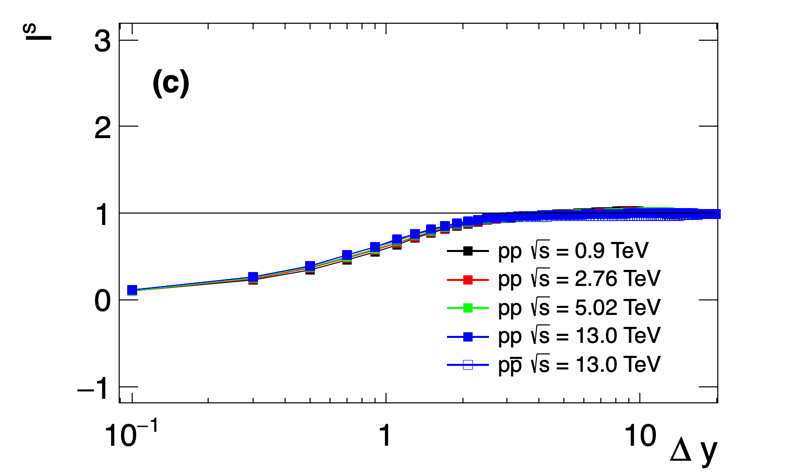}
	\caption{ Cumulative integrals of general  balance functions (a) $B^{-+}$, (b) $B^{+-}$, and (c) $B^{\rm s}$, shown in Fig.~\ref{fig:BFvsy}. }
	\label{fig:PrattIntegralVsY}
\end{figure}
More importantly, these BFs do not integrate to unity across the full rapidity acceptance considered. One finds indeed, as shown in Fig.~\ref{fig:PrattIntegralVsY}, that cumulative integrals of  $B^{-+}(\Delta y,\Delta\varphi)$ and $B^{+-}(\Delta y,\Delta\varphi)$ computed for pp collisions converge towards $-1$ and 3, respectively,   across the full range of the simulations, whereas, by contrast, cumulative integrals of $B^{-+}(\Delta y,\Delta\varphi)$ and $B^{+-}(\Delta y,\Delta\varphi)$ converge to  unity for $\rm p\bar p$ collisions. Clearly, the presence of  long range correlations  shifts  to $1-2=-1$ and $1+2=3$, respectively, the integral of $B^{-+}(\Delta y,\Delta\varphi)$ and $B^{+-}(\Delta y,\Delta\varphi)$. 
One also finds that the cumulative integrals of these functions, for pp collisions,  converge towards 3 and $-1$  very slowly, whereas cumulative integrals  computed for $\rm p\bar p$ converge rather rapidly to unity, i.e. within   $\Delta y<3$. 

\begin{figure}[!ht]
\includegraphics[width=0.32\linewidth,trim={8mm 1mm 26mm 2mm},clip]{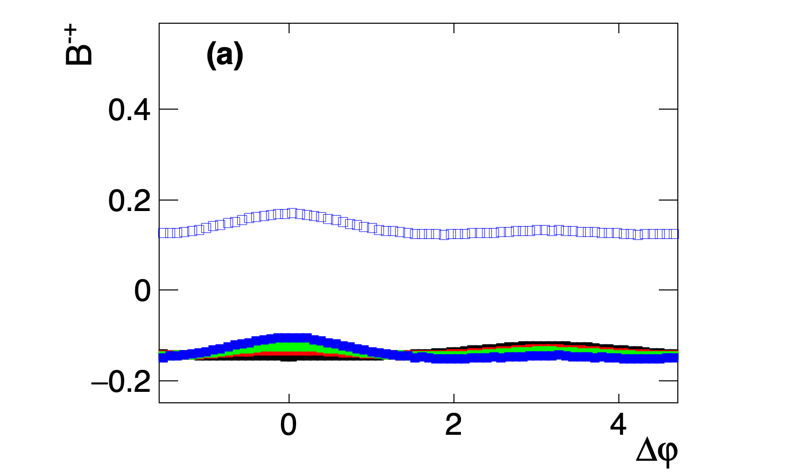}
\includegraphics[width=0.32\linewidth,trim={8mm 1mm 26mm 2mm},clip]{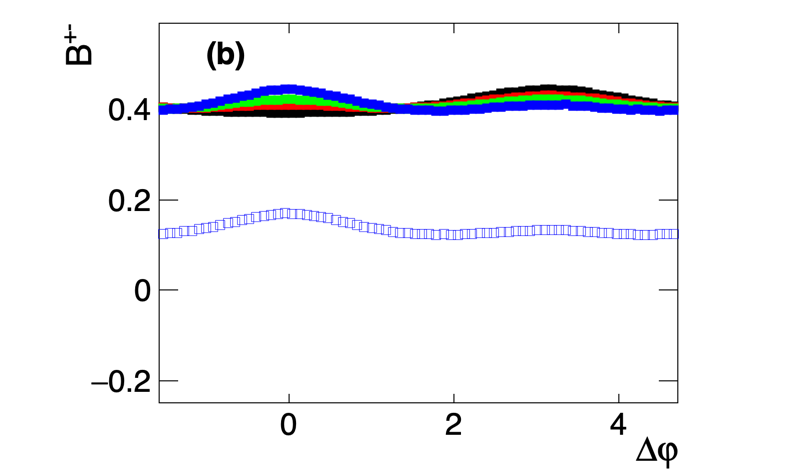}
\includegraphics[width=0.32\linewidth,trim={8mm 1mm 26mm 2mm},clip]{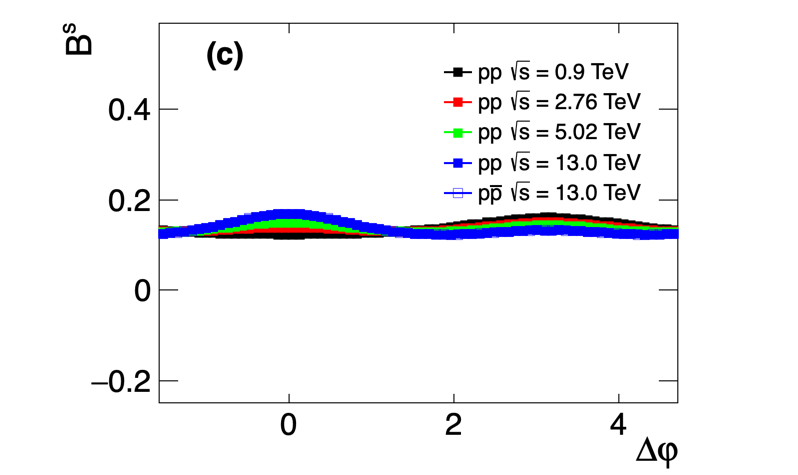}
\caption{ Projections onto the $\Delta \varphi$ axis of balance functions (a) $B^{-+}$, (b) $B^{+-}$, and (c) $B^{\rm s}$ calculated using Eqs.~(\ref{eq:B2vsy1y2},~\ref{eq:B2vsy1y2Bis},~\ref{eq:Bs})  for pp collisions at $\sqrt{s}= 0.9, 2.76, 5.02, 13$ TeV and $\rm p\bar p$ collisions at $\sqrt{s}= 13$ TeV simulated with PYTHIA8. }
\label{fig:BFvsPhi}
\end{figure}

It is  informative to examine  the  $\Delta\varphi$ projections of  $B^{-+}(\Delta y,\Delta\varphi)$ and $B^{+-}(\Delta y,\Delta\varphi)$ shown in Fig.~\ref{fig:BFvsPhi}. 
One finds these have magnitudes very much influenced by the long range correlations seen in Fig.~\ref{fig:BFvsy}. 
Additionally note that the $\Delta \varphi$ projections also evolve significantly with $\sqrt{s}$. At $\sqrt{s}=1$ TeV, $B^{-+}(\Delta y,\Delta\varphi)$ and $B^{+-}(\Delta y,\Delta\varphi)$ feature a maximum value on the away-side, $\Delta\varphi=\pi$, presumably associated with back-to-back momentum conservation in the transverse plane. The amplitude of the away-side, however, is seen to progressively decrease with increasing  $\sqrt{s}$ and eventually leading to a near-side maximum at $\sqrt{s}=13$ TeV in both pp and $\rm p\bar p$ collisions. 
Except for $\rm p\bar p$ collisions, where total net charge of the system vanishes, huge differences are seen between the amplitude of $B^{-+}(\Delta y,\Delta\varphi)$ and $B^{+-}(\Delta y,\Delta\varphi)$, as a result of the long range correlations due to not vanishing total net charge. Clearly, these shifts in amplitude are of limited interest and can only obscure the interpretation of measured balance functions.  It is thus highly desirable to eliminate contributions from the total net charge on long range correlations in the calculation of BFs.

A seemingly natural method  to suppress the long range correlation contribution to BFs is found by considering the arithmetic average $B^{\rm s}(\Delta y,\Delta\varphi)$, as seen in the right-most panel of Fig.~\ref{fig:Pythia:PrattB2in2D}, which displays $B^{\rm s}(\Delta y,\Delta\varphi)$ for pp collisions at $\sqrt{s}=13$ TeV, as well as in Figs.~\ref{fig:BFvsy} and \ref{fig:BFvsPhi}, which show projections onto $\Delta y$ and $\Delta \varphi$ of $B^s$ computed from   $B^{-+}(\Delta y,\Delta\varphi)$ and $B^{+-}(\Delta y,\Delta\varphi)$ displayed in the left and central panels of these figures. One finds that  the ``overshoot" and ``undershoot" of the  long-range components of $B^{-+}(\Delta y,\Delta\varphi)$ and $B^{+-}(\Delta y,\Delta\varphi)$ essentially cancel and yield functions $B^{\rm s}$ that nearly vanish outside of the central rapidity region.  The $\Delta y$ projections of $B^{\rm s}$ are nearly (although not completely) vanishing outside of the range $-5 < y<5$ and the $\Delta \varphi$ projections of $B^{\rm s}$ for pp and $\rm p\bar p$ collisions, both at $\sqrt{s}=13$ TeV, are essentially identical. Parenthetically, also note that the $\sqrt{s}$ evolution of these projections clearly show that charge balancing evolves considerably, according to PYTHIA8, from $\sqrt{s}=0.9$ to 13 TeV, with a maximum (peak) shifting from $\Delta \varphi=\pi$ to $\Delta \varphi=0$.
Finally, integrals of $B^{\rm s}$ computed for pp collisions  for all considered beam energies, as shown in Fig.~\ref{fig:PrattIntegralVsY}, rapidly converge to unity, i.e, within a range less than $\Delta y\sim 3$ as for $\rm p\bar p$ collisions. Using the average $B^{\rm s}$ thus seems a straightforward method  to carry out measurements of balance functions,  and their integrals, that yield the expected behavior and convergence to unity.

Individual measurements of $B^{-+}(\Delta y,\Delta\varphi)$ and  $B^{+-}(\Delta y,\Delta\varphi)$ nonetheless  remain desirable. Experimentally, either of these may not be accessible for practical or technical reasons and a calculation of their average shall then be impossible. It may also be desirable, particularly for species that do not feature $N_{\bar\alpha}/N_{\alpha}=1$, to explicitly compare the two functions. Additionally, one can also verify that species specific (e.g., kaons vs. pions) BFs based on Eqs.~(\ref{eq:B2vsy1y2},~\ref{eq:B2vsy1y2Bis}) do  not, in general, integrate to unity, for systems with non-vanishing net-charge. Addressing the non-vanishing net-charge of a system in measurements of $B^{-+}(\Delta y,\Delta\varphi)$ and  $B^{+-}(\Delta y,\Delta\varphi)$ is thus required. 
  
\section{Unified Balance Functions}
\label{sec:UBF}

Accounting for the non-vanishing net-charge of a colliding system, in computations of balance functions, is readily accomplished by adding single particle densities to the conditional densities considered. 
Indeed, as shown in Ref.~\cite{Pruneau:2022gvt}, it suffices  to include the difference $\rho_1^{\alpha}(y) - \rho_1^{\bar\alpha}(y)$ in the definition of  $B^{\alpha|\bar\beta}(y_1|y_2)$ and $B^{\bar\alpha|\beta}(y_1|y_2)$. This is best  accomplished with the introduction of  {\bf associated particle densities}, denoted  $A_2^{\alpha|\beta}(y_1|y_2)$, and defined according to
\begin{align}
\label{eq:A2}
      A_2^{\alpha|\beta}(y_1|y_2) &\equiv  \frac{C_2^{\alpha\beta}(y_1|y_2)}{\rho_1^{\beta}(y_2)} =  \frac{\rho_2^{\alpha\beta}(y_1,y_2)}{\rho_1^{\beta}(y_2)}   
 - \rho_1^{\alpha}(y_1),
\end{align}
where $C_2^{\alpha\beta}(y_1,y_2)=\rho_2^{\alpha\beta}(y_1,y_2)- \rho_1^{\alpha}(y_1)\rho_1^{\beta}(y_2)$
is a differential two-particle cumulant (correlation function) between particles of species $\alpha$ and $\beta$ emitted at 
rapidities $y_1$ and $y_2$. This function identically vanishes in the absence of particle correlations. Balance functions that  automatically account for a system's non-vanishing net-charge are then written
\begin{align}
\label{eq:B2GenAlphaBarBeta}
      B^{\alpha|\bar\beta}(y_1|y_2) &= A_2^{\alpha|\bar\beta}(y_1|y_2)- 
      A_2^{\bar\alpha|\bar\beta}(y_1|y_2), \\ 
\label{eq:B2GenBarAlphaBeta}
      B^{\bar\alpha|\beta}(y_1|y_2) &= A_2^{\bar\alpha|\beta}(y_1|y_2)- 
      A_2^{\alpha|\beta}(y_1|y_2),
  \end{align}
  and one easily verifies that  Eqs.~(\ref{eq:B2GenAlphaBarBeta},~\ref{eq:B2GenBarAlphaBeta}) integrate to unity in the full acceptance limit, even in the presence of non-vanishing net-charge. Note that   the balance functions $B^{\alpha|\bar\beta}(y_1|y_2)$ and $B^{\bar\alpha|\beta}(y_1|y_2)$ are not positive definite. Indeed, they  may be negative or null across some portions of the acceptance. As such, neither $A_2^{\alpha|\beta}(y_1|y_2)$ nor $B^{\alpha|\beta}(y_1|y_2)$  can  be considered true single particle densities. As already pointed out in Ref.~\cite{Pruneau:2022gvt}, it should be additionally noted that the shape and strength of $A_2^{\alpha|\beta}(y_1|y_2)$ and thus $B^{\alpha|\beta}(y_1|y_2)$
may depend strongly  on $y_2$. For instance, at rapidity $y_2$ near the beam rapidity $y_{\rm b}$, one expects the particle production to be largely dominated by the fragmentation of the beam components whereas at central rapidity ($y\approx 0$
in collider mode), particle production is determined by large $\sqrt{s}$ $\rm q\bar q$ or $gg$  processes. The widths and shapes of BFs are thus indeed expected to vary appreciably with the selected rapidity $y_2$.   Averaging over $y_2$ in a finite measurement acceptance, one gets   bounded balance functions valid for non-vanishing net-charge: 
\begin{align}
\label{eq:B2alphaBarBetay1y2NoQ-C}
   B^{\alpha\bar\beta}(y_1,y_2|y_0) &= 
   \frac{1}{\langle N_1^{\bar\beta}\rangle}\left[ C_2^{\alpha\bar\beta}(y_1,y_2) - C_2^{\bar\alpha\bar\beta}(y_1,y_2) \right]\\
\label{eq:B2BarAlphaBetay1y2NoQ-C}
   B^{\bar\alpha\beta}(y_1,y_2|y_0) &=    \frac{1}{\langle N_1^{\beta}\rangle}\left[ C_2^{\bar\alpha\beta}(y_1,y_2) - C_2^{\alpha\beta}(y_1,y_2) \right].
\end{align}
These two functions, known as UBF, are applicable to same, $\alpha=\beta$, or mixed, $\alpha\ne \beta$,  particle species, each carrying a single unit of charge\footnote{See \cite{Pruneau:2022gvt} for a discussion BFs involving multi-unit charge carriers. }. 

\begin{figure}[!ht]
	\includegraphics[width=0.32\linewidth,trim={8mm 11mm 24mm 3mm},clip]{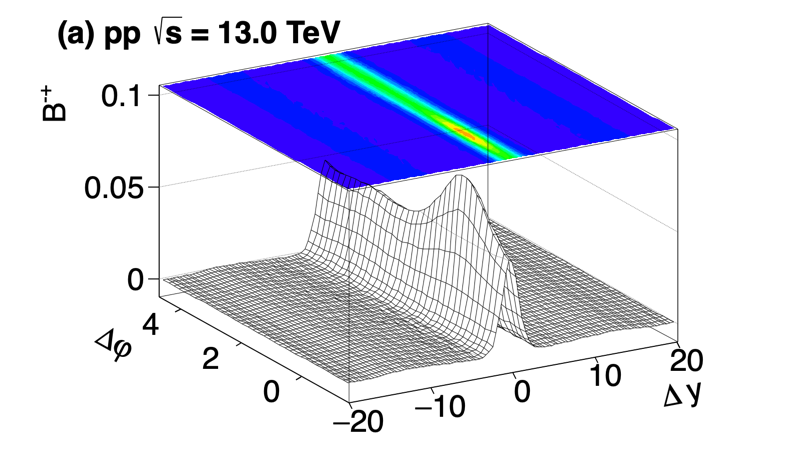}
	\includegraphics[width=0.32\linewidth,trim={8mm 11mm 24mm 3mm},clip]{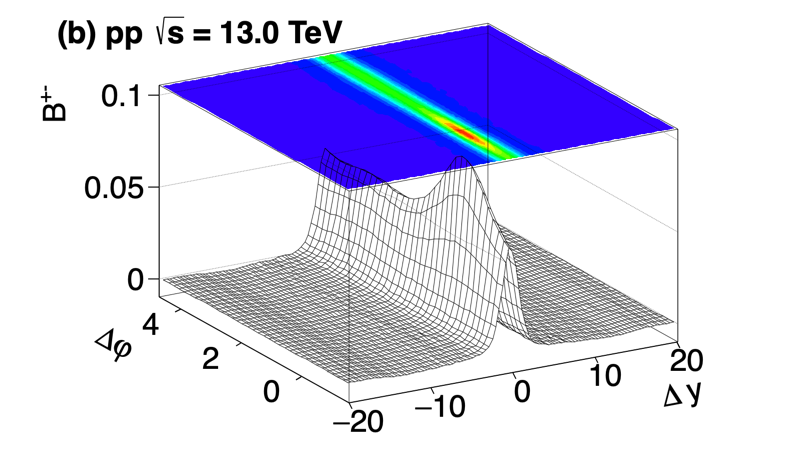}
	\includegraphics[width=0.32\linewidth,trim={8mm 11mm 24mm 3mm},clip]{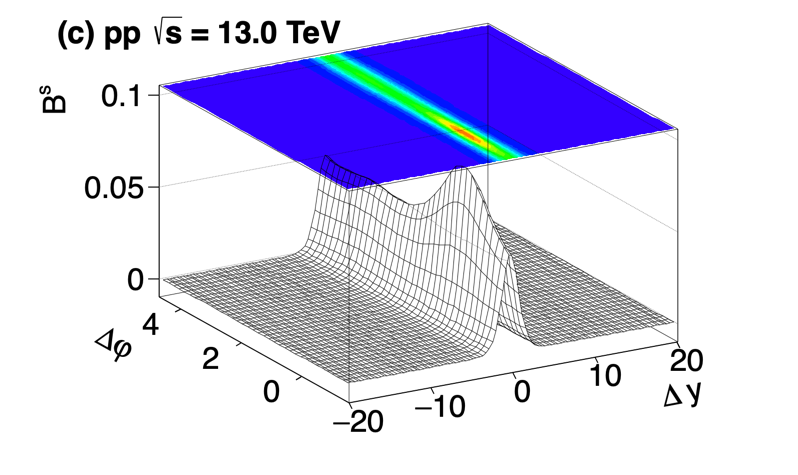}
	\caption{ Unified balance functions (a) $B^{-+}$, (b) $B^{+-}$, and (c) $B^{\rm s}$  calculated using Eqs.~(\ref{eq:B2alphaBarBetay1y2NoQ-C},~\ref{eq:B2BarAlphaBetay1y2NoQ-C},~\ref{eq:Bs})  for pp collisions at $\sqrt{s}= 13$ TeV simulated with PYTHIA8.} 
	\label{fig:Pythia:B2in2D}
\end{figure}

\begin{figure}[!ht]
\includegraphics[width=0.32\linewidth,trim={8mm 1mm 26mm 5mm},clip]{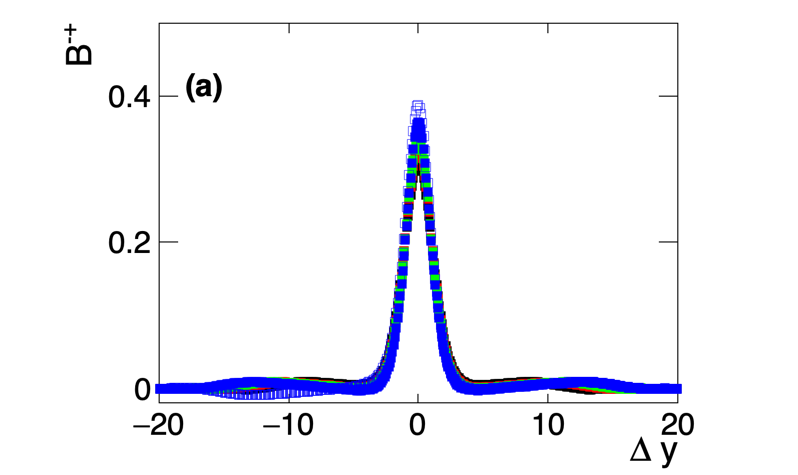}
\includegraphics[width=0.32\linewidth,trim={8mm 1mm 26mm 5mm},clip]{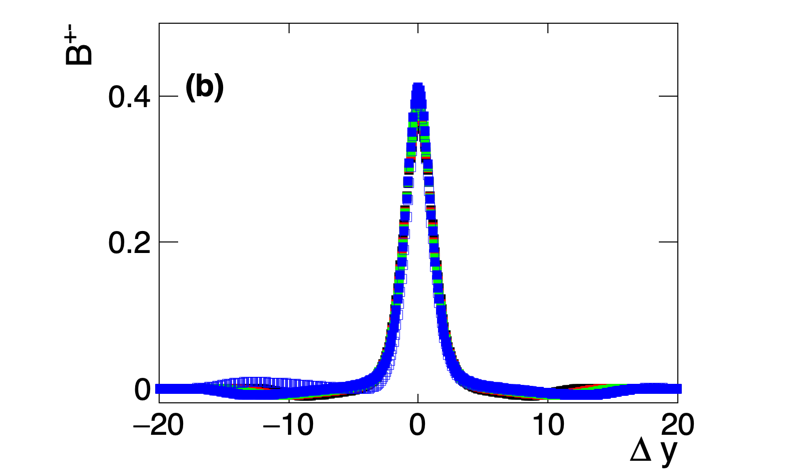}
\includegraphics[width=0.32\linewidth,trim={8mm 1mm 26mm 5mm},clip]{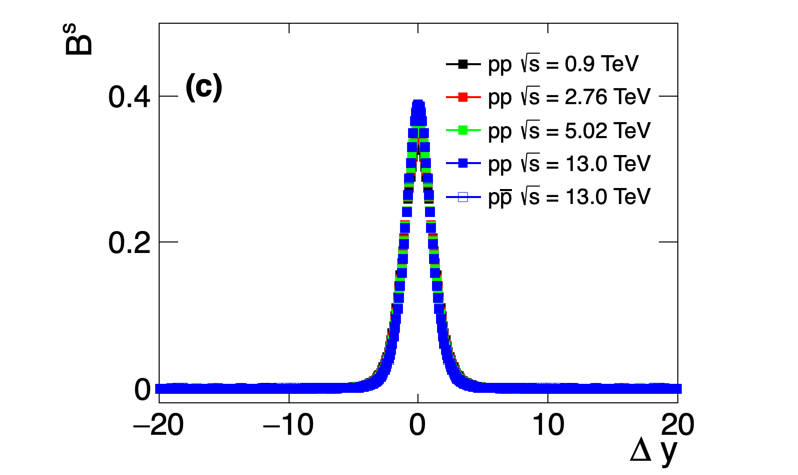}
\caption{ Projections onto the $\Delta y$ axis of UBFs (a) $B^{-+}$, (b) $B^{+-}$  and (c) $B^{\rm s}$ calculated using Eqs.~(\ref{eq:B2alphaBarBetay1y2NoQ-C},~\ref{eq:B2BarAlphaBetay1y2NoQ-C},~\ref{eq:Bs}) for pp collisions at $\sqrt{s}= 1, 2.76, 5.02$ and 13 TeV, as
 well as $\rm p\bar{p}$ collisions at $\sqrt{s}= 13.0$ TeV simulated  with PYTHIA8.}
\label{fig:UBFontoY}
\end{figure}

\begin{figure}[!ht]
\includegraphics[width=0.32\linewidth,trim={8mm 1mm 26mm 1mm},clip]{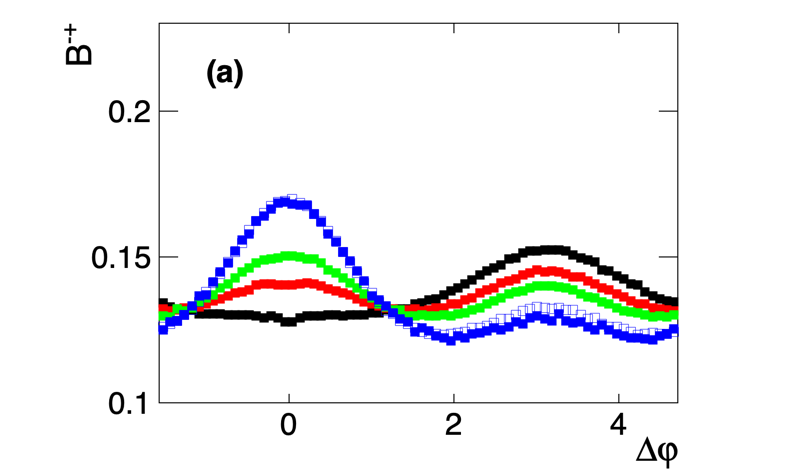}
\includegraphics[width=0.32\linewidth,trim={8mm 1mm 26mm 1mm},clip]{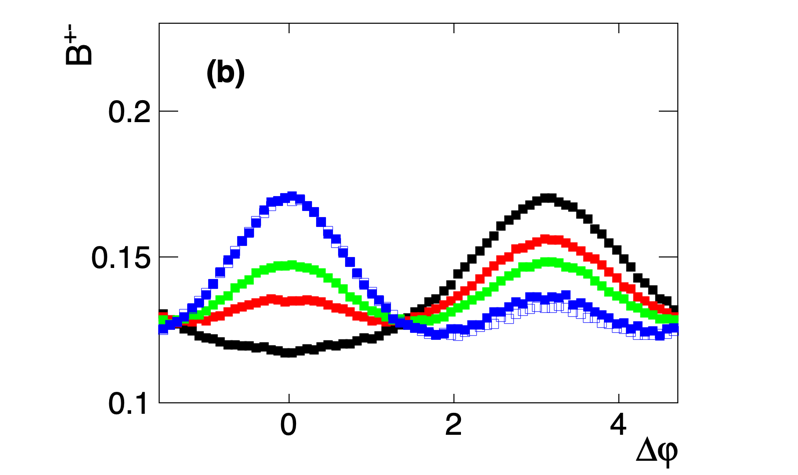}
\includegraphics[width=0.32\linewidth,trim={8mm 1mm 26mm 1mm},clip]{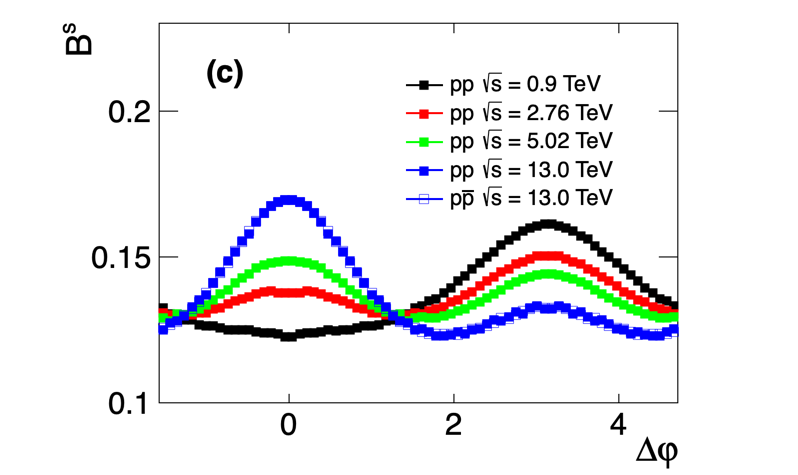}
\caption{Projections onto the $\Delta \varphi$ axis of UBFs (a) $B^{-+}$, (b) $B^{+-}$  and (c) $B^{\rm s}$ calculated using Eqs.~(\ref{eq:B2alphaBarBetay1y2NoQ-C},~\ref{eq:B2BarAlphaBetay1y2NoQ-C},~\ref{eq:Bs}) for pp collisions at $\sqrt{s}= 1, 2.76, 5.02$ and 13 TeV, as
 well as $\rm p\bar{p}$ collisions at $\sqrt{s}= 13.0$ TeV simulated  with PYTHIA8.}
\label{fig:UBFontoPhi}
\end{figure}

\begin{figure}[!ht]
	\includegraphics[width=0.32\linewidth,trim={8mm 1mm 26mm 5mm},clip]{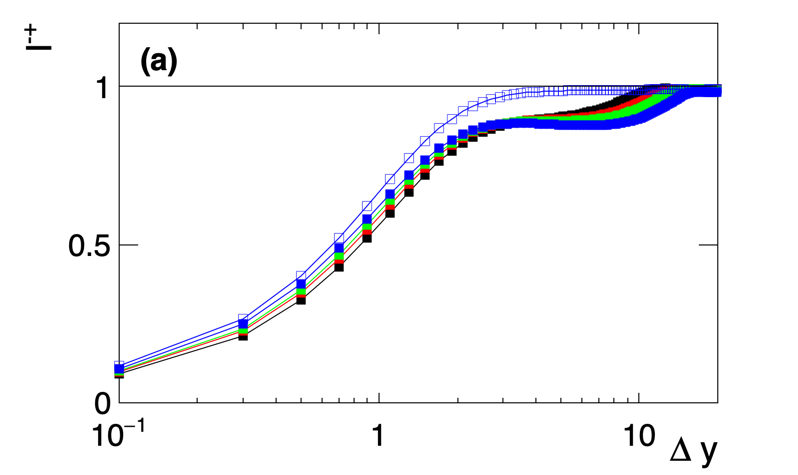}
	\includegraphics[width=0.32\linewidth,trim={8mm 1mm 26mm 5mm},clip]{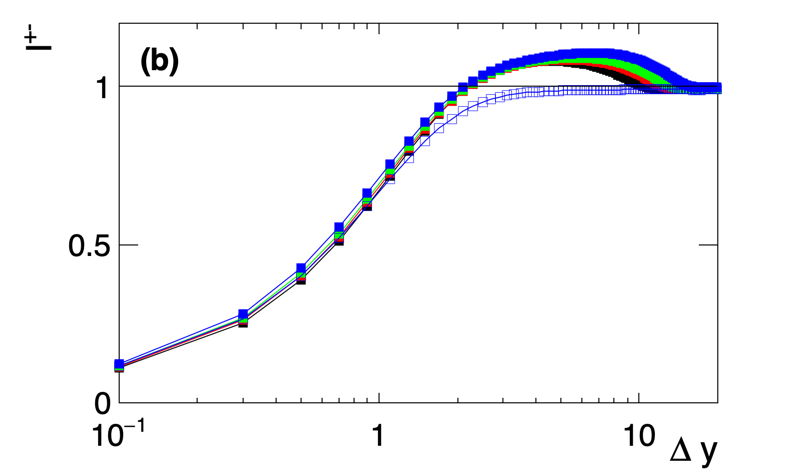}
	\includegraphics[width=0.32\linewidth,trim={8mm 1mm 26mm 5mm},clip]{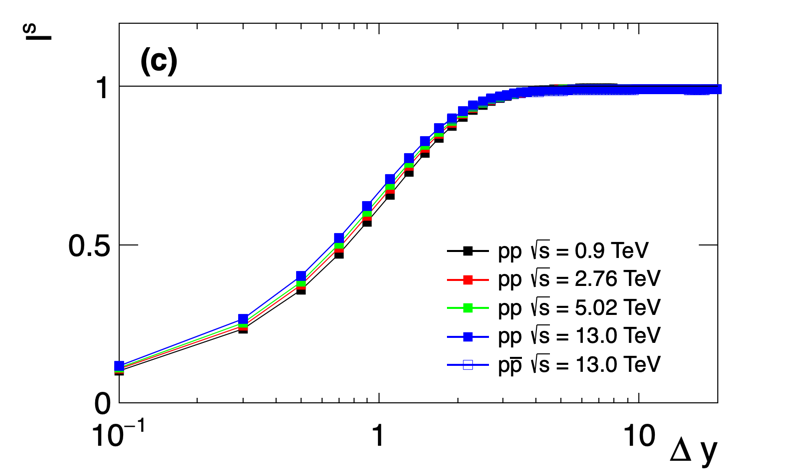}
\caption{ Cumulative integrals, $I(\Delta y)$, of UBF calculated with Eqs.~(\ref{eq:B2alphaBarBetay1y2NoQ-C},~\ref{eq:B2BarAlphaBetay1y2NoQ-C},~\ref{eq:Bs}), plotted as a function of the acceptance $|\Delta y|$  for pp collisions at $\sqrt{s}= 1, 2.76, 5.02$ and 13 TeV as
 well as $\rm p\bar{p}$ collisions $\sqrt{s}= 13.0$ TeV simulated with PYTHIA8.}
\label{fig:UnifiedI2vsDeltaY}
\end{figure}

Figure~\ref{fig:Pythia:B2in2D} displays $B^{+-}(y_1,y_2|y_0)$ and $B^{-+}(y_1,y_2|y_0)$, computed with Eqs.~(\ref{eq:B2alphaBarBetay1y2NoQ-C},~\ref{eq:B2BarAlphaBetay1y2NoQ-C}) for $\alpha=\beta=+$ and $\bar\alpha=\bar\beta=-$, and their  average, $B^{\rm s}$, obtained for pp collisions at $\sqrt{s}=$ 13 TeV simulated with PYTHIA8. 
Projections of these UBFs and those obtained for pp collisions at $\sqrt{s}=0.90$, 2.76, 5.02 TeV and $\rm p\bar{p}$ collisions at 13 TeV are shown in Figs.~\ref{fig:UBFontoY} and \ref{fig:UBFontoPhi}.  One readily verifies, 
based  on Fig.~\ref{fig:Pythia:B2in2D}, and the longitudinal projections shown in Fig.~\ref{fig:UBFontoY}, that UBFs     feature nearly vanishing long range components, in stark contrast to GBFs computed with Eqs.~(\ref{eq:B2vsy1y2},~\ref{eq:B2vsy1y2Bis}).  One finds indeed that the UBFs are dominated by their central rapidity peak and feature very small long range components beyond $|y|>4$ or so. 
Note, in particular,  that  $B^{-+}(y_1,y_2|y_0)$, $B^{+-}(y_1,y_2|y_0)$ have only slight 
overshoots and undershoots, respectively, for pp collisions and a small undershoot and overshoot, respectively, for $\rm p\bar p$ collisions. These small positive and negative long-range excesses vanish identically in the average correlator $B^{\rm s}$. The effect of the non-vanishing remaining long-range components is best seen in cumulative integrals of the BFs shown in Fig.~\ref{fig:UnifiedI2vsDeltaY}. Indeed note that for pp collisions, the cumulative integral of  $B^{\alpha\bar\beta}(y_1,y_2|y_0)$ quickly rises to near unity at $y=3$, but subsequently converge rather slowly towards unity near twice the beam rapidity. In the case of $B^{\bar\alpha\beta}(y_1,y_2|y_0)$, one finds that  the rise first exceeds unity near $y=3$ to eventually converge back, very slowly, to unity at twice the beam rapidity. Such effects cancel in the average $B^{\rm s}$. We thus conclude that UBFs $B^{\alpha\bar\beta}(y_1,y_2|y_0)$ and $B^{\bar\alpha\beta}(y_1,y_2|y_0)$, defined by Eqs.~(\ref{eq:B2alphaBarBetay1y2NoQ-C},~\ref{eq:B2BarAlphaBetay1y2NoQ-C}), carry much smaller effects from long-range correlations due to non vanishing net charge than  the basic GBFs computed according to Eqs.~(\ref{eq:B2vsy1y2},~\ref{eq:B2vsy1y2Bis}), and, as such, may be considered much more acceptable measures of balance functions if perfect precision is not required. However, the sum $B^{\rm s}$ does not involve these effects and its use should thus be preferred whenever possible, particularly, if the magnitude of the integral of these function is of prime interest. 

Shifting the focus onto Fig.~\ref{fig:UBFontoPhi}, we once again note  a sizable change in the $\Delta \varphi$ dependence of  $B^{\rm s}$ with collision energy.  At the lowest energy considered, $\sqrt{s}=0.9$~TeV,  PYTHIA yields a BF with a slight excess on the away-side ($\Delta\varphi=\pi$). The excess is found to progressively decrease with rising   $\sqrt{s}$, yielding near- and away-side  ``peaks" of approximately equal height at $\sqrt{s}=5.02$~TeV, and a significant near-side excess at $\sqrt{s}=13.0$~TeV in both pp and $\rm p\bar p$ collisions.  This evolution with  $\sqrt{s}$ is likely due to  the fast rise of the jet cross section from $\sqrt{s}=1$ to 13 TeV. Jets are by definition nearly charge neutral and charge correlations from one jet to another are expected to be weak given particle production within a jet is essentially limited to the cone of the jet. As the jet cross section rises, near-side charged correlations, corresponding to intra-jet correlations, progressively dominate and the maximum of BFs then shifts from the away-side to the near-side. 

\section{Impact of limited acceptance on UBF measurements}
\label{sec:impact}

Experimentally, achieving efficient detector coverage  and good charged particle momentum resolution  at  rapidities in excess of $y=4$ is   rather challenging.  Indeed, detectors in operation, e.g., at RHIC and the LHC, have rapidity coverage typically limited to central rapidities. Currently, ALICE has transverse momentum measurement capabilities for $|y|<1$ with a low $p_{\rm T}$ threshold of about 0.15 GeV/$c$,  while CMS and ATLAS have acceptances up to 3 or 4 units of pseudorapidity but feature poor charged particle track purity at $p_{\rm T} < 0.5 $ GeV/$c$. However, technologies envisioned  for ALICE 3 \cite{ALICE:2803563} might extend the experimental acceptance to both lower momenta and larger rapidities but are unlikely to go much beyond $y=4$. We thus explore in somewhat more detail the quality of measurements that can be carried with UBFs using practical  values for $p_{\rm T}$ and $y$ acceptances. 

\begin{figure}[!ht]
\includegraphics[width=0.32\linewidth,trim={8mm 1mm 26mm 5mm},clip]{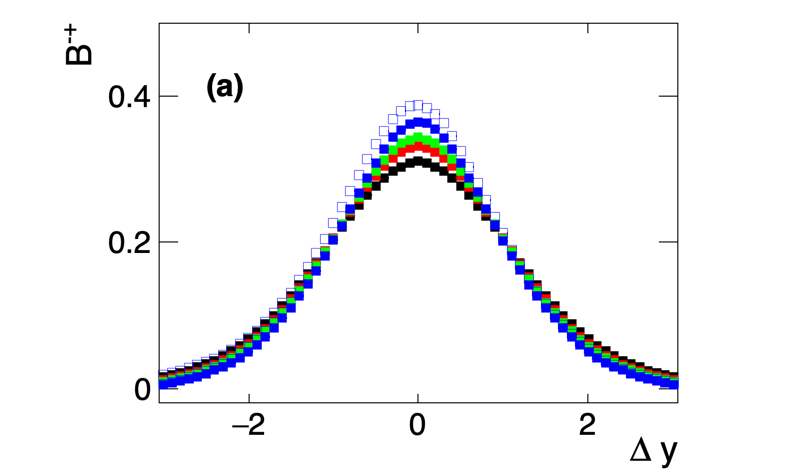}
\includegraphics[width=0.32\linewidth,trim={8mm 1mm 26mm 5mm},clip]{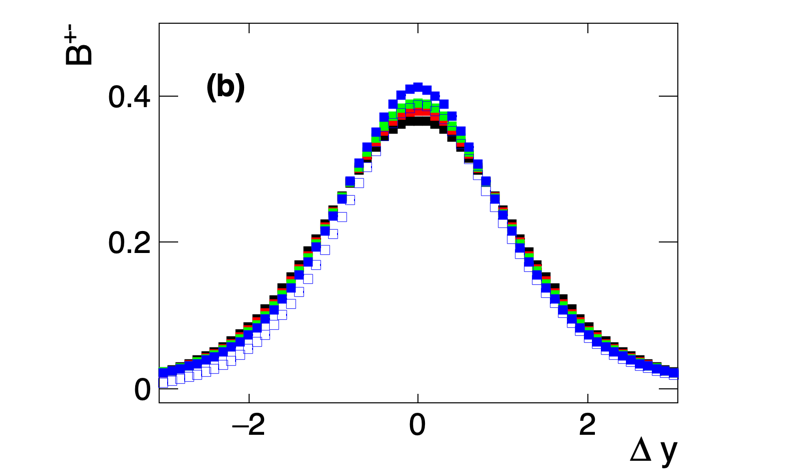}
\includegraphics[width=0.32\linewidth,trim={8mm 1mm 26mm 5mm},clip]{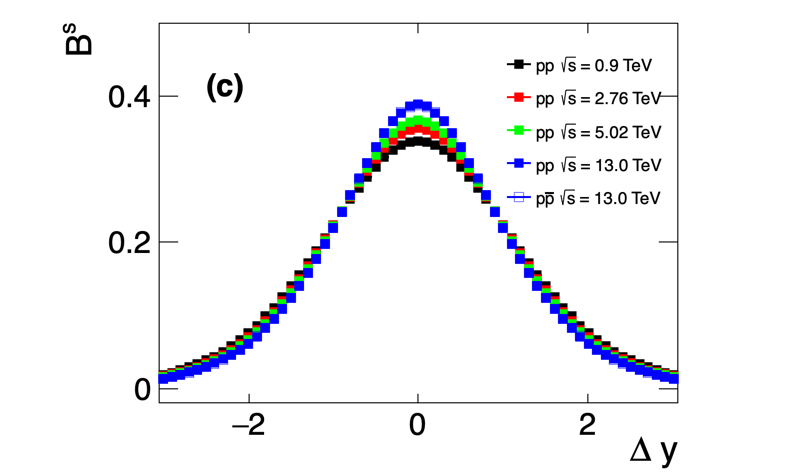}
\caption{ Projections onto the $\Delta y$ axis of UBFs (a) $B^{-+}$, (b) $B^{+-}$, and (c) $B^{\rm s}$ calculated 
using Eqs.~(\ref{eq:B2alphaBarBetay1y2NoQ-C},~\ref{eq:B2BarAlphaBetay1y2NoQ-C},~\ref{eq:Bs}) for pp collisions at $\sqrt{s}= 1, 2.76, 5.02$ and 13 TeV, as  well as $\rm p\bar{p}$ collisions at $\sqrt{s}= 13.0$ TeV simulated  with PYTHIA8.}
\label{fig:B2ontoDeltaY-HR}
\end{figure}

Let us first examine in further detail UBFs obtained with a very wide (perfect) acceptance. 
Figure~\ref{fig:B2ontoDeltaY-HR} displays UBFs, computed within the full  acceptance $y_0=10$, in the range of interest, $-3 < \Delta y < 3$, for several beam energies. 
One finds that PYTHIA8 produces UBFs with  magnitudes and widths that feature a modest dependence on $\sqrt{s}$. Note, in particular, that $B^{-+}(y_1,y_2|y_0)$ and $B^{+-}(y_1,y_2|y_0)$
exhibit hierarchically ordered amplitudes, i.e., amplitudes that progressively grow with increasing $\sqrt{s}$ for rapidity separations $|\Delta y| < 1$ and with decreasing $\sqrt{s}$ for rapidity separation $2 < |\Delta y| < 3$. Amplitudes of UBFs $B^{+-}$ for pp are larger than their counterparts $B^{-+}$ for all collision energies while keeping their  hierarchical order. Amplitude of the pp at 13 TeV UBF moves from lower values than the $\rm p \bar p$ $B^{-+}$ to higher values than the $\rm p \bar p$ $B^{+-}$ while both UBFs for the $\rm p \bar p$ system are basically identical.  
The hierarchical order according to the system energy is exhibited in the average $B^{\rm s}$ where the BFs corresponding to the pp and $\rm p \bar p$ at 13 TeV systems overlap as expected from such ordering. 

\begin{figure}[!ht]
	\includegraphics[width=0.48\linewidth,trim={8mm 1mm 8mm 1mm},clip]{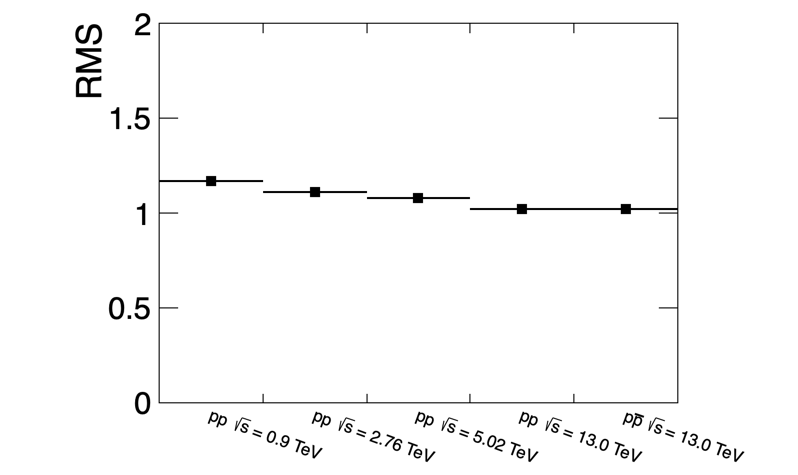}
\caption{ RMS widths of UBFs $B^s$ computed at selected values of $\sqrt{s}$ for  pp collisions  and  at  $\sqrt{s}= 13.0$ TeV for $\rm p\bar{p}$ collisions simulated  with PYTHIA8. }
\label{fig:UBF-WidthsVsSqrtS}
 \end{figure}

The UBFs shown in Fig.~\ref{fig:B2ontoDeltaY-HR} are found to be well described by Gaussian probability density functions (PDF). 
One finds, in particular, that the behavior of the $B^{\rm s}$ amplitude with $\sqrt{s}$ is accompanied by a monotonic decrease  of their rms width, as illustrated in  Fig.~\ref{fig:UBF-WidthsVsSqrtS}, which shows the rms widths of UBFs $B^{\rm s}$ for pp collisions at $\sqrt{s}=$ 0.9, 2.76, 5.02, and 13.0 TeV and for $\rm p\bar p$ collisions at $\sqrt{s}=13.0$ TeV. One   notes, additionally, that the rms widths for pp and $\rm p\bar p$ collisions at $\sqrt{s}=13.0$ TeV are identical within statistical uncertainty. 
This suggests that charged particle balancing, at least in the context of the PYTHIA8 model,  is independent of the incoming beam particle species in this high-energy collision regime. Additionally note that the slight narrowing of the UBFs with increasing $\sqrt{s}$ is expected on general grounds. The average transverse momentum $\langle p_{\rm T}\rangle$ increases monotonically with $\sqrt{s}$. Correlated particles resulting from decays, jet fragmentation, etc, thus have a tendency to be kinematically focused, i.e, emitted at smaller rapidity and azimuthal angle  differences with increasing $\sqrt{s}$. The rate of this rms narrowing with increasing $\sqrt{s}$ and the rise of the UBF amplitude, along with changes of the detailed shape of the UBFs, are evidently candidates for observables capable of discriminating the performance of particle production models.  However, it should be clear that the results shown in Fig~\ref{fig:B2ontoDeltaY-HR} were obtained for a perfect acceptance in rapidity, azimuth, and transverse momentum. It is legimate to expect that measurements biases may occur with reduced acceptance in (pseudo)rapidty and transverse momentum. We thus explore, in the following, what may be the impact of such reduced acceptances on measurements of UBFs, particularly, their shape (amplitude vs. width) as well as their integrals. 

We next examine the impact of the limited acceptance in rapidity on measurements of UBFs. Our study is based on the  assumption that particle species  can be identified and that their rapidity computed from their momentum. It is clear that further limitations would arise if particles cannot be identified and measurements are thus limited to the pseudorapidity of particles, but such studies are left for further works and detailed studies.

\begin{figure}[!ht]
\includegraphics[width=0.32\linewidth,trim={8mm 1mm 26mm 5mm},clip]{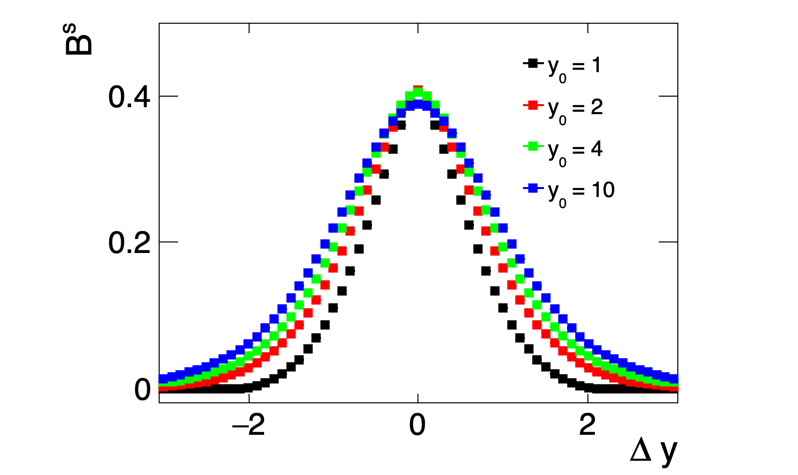} 
\includegraphics[width=0.32\linewidth,trim={8mm 1mm 26mm 5mm},clip]{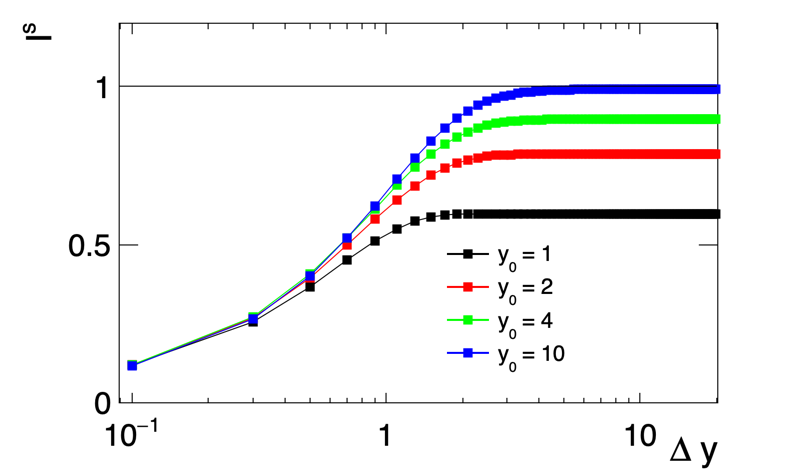}
\includegraphics[width=0.32\linewidth,trim={8mm 1mm 26mm 1mm},clip]{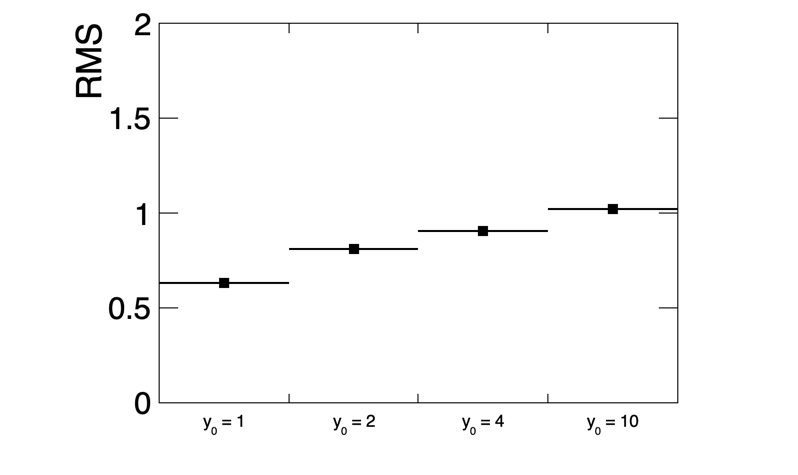}
\caption{(left) UBFs $B^{\rm s}(\Delta y)$ computed for pp collisions at $\sqrt{s}=13$ TeV, for selected longitudinal  acceptances; (center) cumulative integrals and (right) RMS widths of UBFs shown in the left panel.}
\label{fig:yDependance}
\end{figure}

The left panel of Fig.~\ref{fig:yDependance} compares projections of UBFs $B^{\rm s}$  computed  in longitudinal acceptances $y_0=1, 2, 4$ and 10. The integrals of these UBFs are shown in the central panel and their rms widths in the right panel of Fig.~\ref{fig:yDependance}.  
One observes that the UBF obtained with $y_0=4$ is partially clipped, relative to that computed with the full acceptance $y_0=10$, and its rms width is thus slightly reduced. One additionally finds that roughly 10\% of the charge balance is lost.  The impact of the the reduction in acceptance obtained with values $y_0=2$ and $y_0=1$ is significantly more dramatic: larger reductions of the rms width of these UBF are found relative to the UBF obtained $y_0=4$. One also notes that the reduction of the integrals of these UBFs is severe,  featuring a 45\% loss of balancing charges for $y_0=1$. Clearly, measurements of UBFs should be carried out in as wide a rapidity acceptance to avoid dramatic modification of their shape (width) and loss of charge balance. 

\begin{figure}[!ht]
\includegraphics[width=0.32\linewidth,trim={8mm 1mm 16mm 5mm},clip]{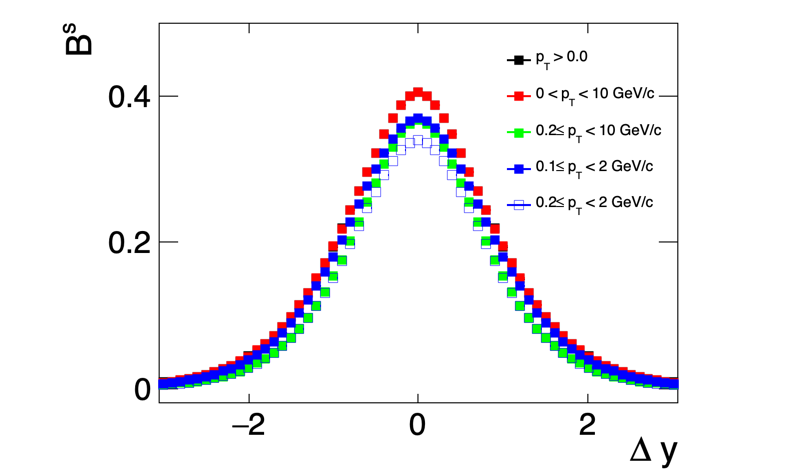}
\includegraphics[width=0.32\linewidth,trim={8mm 1mm 16mm 5mm},clip]{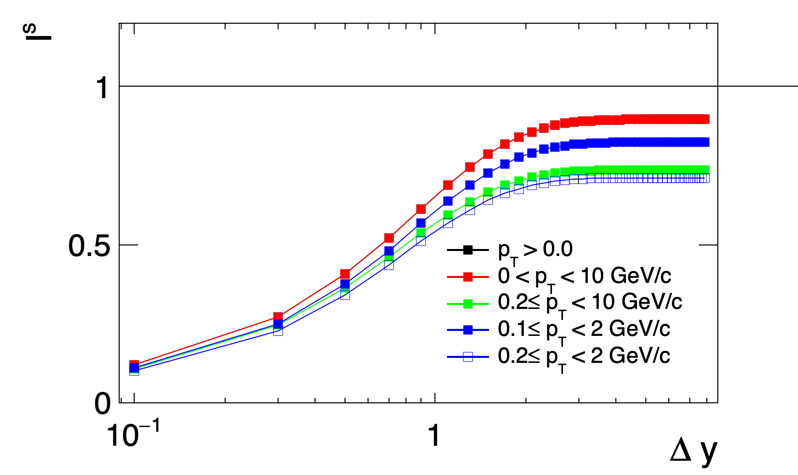}
\includegraphics[width=0.32\linewidth,trim={8mm 1mm 16mm 1mm},clip]{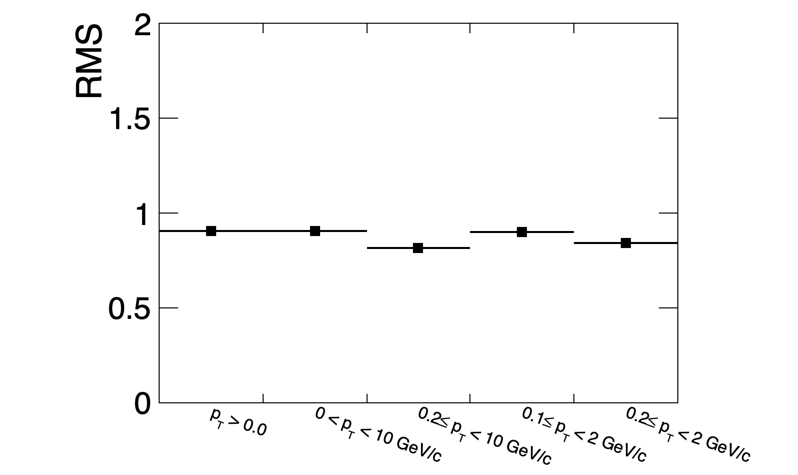}
\caption{ (left) UBFs $B^{\rm s}(\Delta y)$ computed for pp collisions at $\sqrt{s}=13$ TeV within an experimental acceptance of $y_0=4$ and selected $p_{\rm T}$ ranges; (center) cumulative integrals and (right) RMS widths of UBFs shown in the left panel.}
\label{fig:Pythia:pTdependance}
\end{figure}

With the hope that an acceptance $y_0=4$ may be achievable within future experiments, we next consider the impact of a reduction in transverse momentum  within this longitudinal acceptance. Figure~\ref{fig:Pythia:pTdependance} displays projections, integrals, and rms widths of UBFs obtained with three selected $p_{\rm T}$ ranges compared to that achieved with an ideal range, $p_{\rm T}>0$.
Based on these simulations with PYTHIA8, we find that  reducing the $p_{\rm T}$ acceptance has a  relatively small impact on the shape and rms width of the distributions but may impact their integral significantly. 
The charge balance integral is most affected by particle losses at low momentum while losses at high $p_{\rm T}$ have only a minor impact on the UBF integrals. Indeed, while reducing the acceptance to $0<p_{\rm T}<10$, only a negligible    loss of charge balance of $<$1\% relative to the $p_{\rm T}>0$  reference is observed. However, reducing the maximum of the range to $p_{\rm T}=2.0$ GeV/$c$ reduces the charge balance by $\sim 7$\%. By contrast, cutting the acceptance by rising the minimum $p_{\rm T}$ to 0.1 and 0.2 GeV/$c$ has a more  significant effect.  With a detection threshold 
of 0.1 GeV/$c$ a charge balance loss of 10\% is incurred, relative to the reference $p_{\rm T}>0$. Cutting the range both from above and below produces the strongest losses, with an integral of 0.83 with   $0.1<p_{\rm T}<2$ GeV/$c$ and  0.71 for  $0.2<p_{\rm T}<2$ GeV/$c$. It is thus clear that designs of future experiments should prioritize a reduction of the detection $p_{\rm }$ threshold in order to optimize the quality of measurements of BFs.  This should also suggest that UBFs studies should also be conducted as function of the particle's transverse momentum. Studies with low-$p_{\rm T}$ ranges might  enable a better understanding of pair production and transport mechanisms, whereas  the structure of jets could be probed based on high-$p_{\rm T}$ ranges.

\section{Summary}
\label{sec:summary} 

We presented a study of general (GBF) and unified (UBF) balance functions based on simulations of pp and $\rm p\bar p$ collisions with  PYTHIA8. We first showed that for collision systems with a non-vanishing net-charges, $Q=2$ in the present case, bound general balance functions $B^{+-}$ and $B^{-+}$, defined by Eqs.~(\ref{eq:B2PrattA},~\ref{eq:B2PrattB}), respectively, contain large long range components that reflect correlations between ``stopped" beam particles and particle pairs created out the vacuum. We showed the integral of $B^{+-}$ and $B^{-+}$ converge to  $1+Q=3$ or $1-Q=-1$, respectively, instead of unity as expected for balanced charge production. We found that GBFs obtained for $\rm p\bar p$ also feature long range components but their contributions change sign below and above $\Delta y=0$ and their integral thus properly converge to unity.  We next showed that UBFs, defined by Eqs.~(\ref{eq:B2alphaBarBetay1y2NoQ-C},~\ref{eq:B2BarAlphaBetay1y2NoQ-C}), largely suppress but do not eliminate the long range correlations seen in GBFs. However, integrals of these UBFs do converge to unity in the full acceptance limit. We additionally showed that averages of GBFs and UBFs, $B^{\rm s}$, completely suppress long range components and feature integrals that properly converge to unity in the large acceptance limit.  Given experiments typically feature narrow acceptances in rapidity (or pseudorapidity) and non-vanishing transverse momentum detection threshold, we next examined the impact of restricting the experimental acceptance both in rapidity and in transverse momentum. We found that narrow longitudinal acceptance significantly impact the shape and width of longitudinal projections of UBFs and their integrals. We additionally found that losses of particles at high-$p_{\rm T}$ have little impact on the shape and integrals of UBFs and  produce small changes of the width and no dramatic loss of charge balance while the larger impact is produced when the low $p_{\rm T}$ detection threshold is raised. As we pointed out in the introduction, integrals of BFs are closely connected to the magnitude of net-charge multiplicity cumulants $\kappa_2$. Conclusions applying to measurements of the  integral of BFs thus also apply to measurements of $\kappa_2$ and caution should consequently be exercised in the interpretation of integral measurements of net charge in A--A collisions. 

Although desirable, correction for pair losses incurred with narrow longitudinal acceptance and finite $p_{\rm T}$ detection threshold were not attempted in the context of this work.
Unified charge balance functions obtained for pp collisions at several distinct values of $\sqrt{s}$ are found to evolve in shape with $\sqrt{s}$. Modifications of the shapes of these balance functions are thus also functions of  $\sqrt{s}$, as are the charge balance losses observed with reduced longitudinal and transverse momentum acceptance. It is thus rather unlikely that reliable and model independent corrections for losses associated with reduced acceptance   
can be devised and implemented. Although not discussed in this work, it can be shown that particle losses associated with instrumental effects (detection efficiency) are  correctable using common techniques and thus have minimal impact on measurements of balance functions. Information modified by the response of a detector can be properly corrected for, but the information falling outside the bounds of  measurements is usually lost. In some cases, such as single particle $p_{\rm T}$ spectra, extrapolation for expected behavior at  $p_{\rm T}\rightarrow 0$ and $p_{\rm T}\rightarrow \infty$ enable estimate of the integral of these spectra. Sadly, in the case of balance of functions, information lost is irremediably lost: it is thus important to design detection devices with as large an acceptance as possible.

\newenvironment{acknowledgement}{\relax}{\relax}
\begin{acknowledgement}
	\section*{Acknowledgements}

	The authors thank Dr. Scott Pratt for insightful discussions and  suggestions. 
	SB acknowledges the support of the Swedish Research Council (VR) and the Knut and Alice Wallenberg Foundation. This work  was also supported in part by the United States Department of Energy, Office of Nuclear Physics (DOE NP), United States of America, under grant No.  DE-FG02-92ER40713. 

\end{acknowledgement}

\bibliography{main}

\end{document}